\documentclass[aps,prl,twocolumn,showpacs,groupedaddress]{revtex4}
\usepackage{color} 
\usepackage{epsfig}
\preprint{TOP INS 2011}
\begin{document}
\bibliographystyle{prsty}

\title{Optical conductivity of Bismuth-based topological insulators} 
\author{P. Di Pietro$^{1}$, F. M. Vitucci$^{1}$, D. Nicoletti$^{2}$, L. Baldassarre$^{3}$,  P. Calvani$^{1}$, R. Cava$^{4}$, Y.S. Hor$^{4}$, U. Schade$^{5}$ and S. Lupi$^{6}$}
\affiliation {$^{1}$ CNR-SPIN and Dipartimento di Fisica, Universit\`a di Roma ``La Sapienza'',
Piazzale A. Moro 2, I-00185 Roma, Italy}
\affiliation {$^{2}$Max Planck Research Department for Structural Dynamics
Center for Free Electron Laser Science \& University of Hamburg, Notkestrasse 85 - 22607 Hamburg, Germany}
\affiliation {$^{3}$Sincrotrone Trieste, Area Science Park, I-34012 Trieste, Italy}
\affiliation {$^{4}$ Department of Chemistry, Princeton University, New Jersey 08544, U.S.A.}
\affiliation{$^{5}$Berliner Elektronenspeicherring-Gesellshaft f\"ur Synchrotronstrahlung m.b.H., Albert-Einstein Strasse 15, D-12489 Berlin, Germany}
\affiliation {$^{6}$ CNR-IOM, and Dipartimento di Fisica, Universit\`a di Roma ``La Sapienza'',
Piazzale A. Moro 2, I-00185 Roma, Italy}

\date{\today}

\begin{abstract} 
The optical  conductivity $\sigma_{1}(\omega)$ and the spectral weight $SW$ of
four topological insulators with increasing chemical compensation (Bi$_{2}$Se$_{3}$, Bi$_{2-x}$Ca$_{x}$Se$_{3}$, Bi$_{2}$Se$_{2}$Te, Bi$_{2}$Te$_{2}$Se) have been measured from 5 to 300 K and from sub-THz to visible frequencies. The effect of compensation is clearly observed in the infrared spectra, through the suppression of an extrinsic Drude term and the appearance of strong absorption peaks, that we assign to electronic transitions among localized states. From the far-infrared spectral weight $SW$ of the most compensated sample (Bi$_2$Te$_2$Se) one can estimate a density of charge-carriers in the order of $10^{17}$/cm$^3$ in good agreement with transport data. Those results demonstrate that the low-energy electrodynamics in single crystals of topological insulators, even at the highest degree of compensation presently achieved, is still affected by extrinsic charge excitations.
\end{abstract}


\maketitle

Topological Insulators (TI) are new quantum materials with an insulating gap in the bulk, of spin-orbit origin, and metallic states at the surface \cite{Hasan1, Kane1, Moore1}. These states are chiral and show dissipation-less transport properties protected from disorder by the time-reversal symmetry. In addition to their fundamental properties, like exotic superconductivity \cite{Fu, Nilsonn} and \textit{axionic} eletromagnetic response \cite{Qi, Essin}, TI have potential applications in quantum computing \cite{Fu1, Kitaev1}, THz detectors \cite{zangh} and spintronic devices \cite{Chen}.  After the discovery of a topological behavior in three-dimensional (3D) Bi$_{x}$Sb$_{1-x}$ \cite{Hsieh}, Bi$_{2}$Se$_3$ recently emerged, thanks to its large bulk insulating gap ($E_G \sim$300 meV), as the best candidate for the study of topological surface states \cite{Zhang}. In fact, Dirac quasi-particles (DQP) related to the metallic surface states have been detected through Angle Resolved Photoemission Spectroscopy (ARPES)  in Bi$_{2}$Se$_{3}$, Bi$_{2}$Te$_{3}$ and in their alloys Bi$_{2}$Se$_{2}$Te and Bi$_{2}$Te$_{2}$Se \cite{Xia, Hsieh}. Investigating the charge transport and cyclotron resonances of DQP has, however, proven to be challenging, because the surface current contribution is usually obscured by the extrinsic bulk carriers response [\onlinecite{Qu, Basov, Butch}]. 

Indeed, as-grown crystals of Bi$_{2}$Se$_{3}$ display a finite density of Se vacancies which act as electron donors. They pin the bulk chemical potential $\mu_{b}$ within the conduction band thus producing, over a wide range of carrier concentrations, extrinsic \textit{n}-type degenerate semiconducting behavior. Se vacancies also affect the low-energy transport properties of those materials \cite{Cava1}, making it difficult to distinguish the intrinsic metallic behavior due to the topological surface state from the extrinsic metallic conduction induced by the Se non-stoichiometry. As a consequence, both transport and optical conductivity experiments \cite{Black, Basov} show a metallic behavior with a  Drude term confined at low frequencies ($\omega<600$ cm$^{-1}$) which mirrors the extrinsic carrier content. Two phonon peaks interacting with the electronic continuum have been observed in the far-IR range near 61 cm$^{-1}$ ($\alpha$ mode) and 133 cm$^{-1}$ ($\beta$ mode) \cite{Basov, Richter}. The bulk insulating gap instead spans between 250-350 meV, depending on the Se vacancy content, in good agreement with theoretical calculations \cite{Zhang1}. 
At variance with Bi$_{2}$Se$_{3}$, single crystals of Bi$_{2}$Te$_{3}$ display $p$-type conductivity related to an excess of Bi atoms acting as acceptor centers \cite{Qu}. These shift $\mu_{b}$ into the valence band so that, like for Bi$_{2}$Se$_3$, an extrinsic Drude term is observed in the far infrared (here below $\omega < 400$ cm$^{-1}$) \cite{Thomas}. 

Motivated by the above observations, different authors adopted specific strategies to reduce the non-stoichiometry-induced bulk carriers in Bi$_{2}$Se$_{3}$ (Bi$_{2}$Te$_{3}$) materials. Hor \textit{et al.} showed that Ca-doping in the Bi site (Bi$_{2-x}$Ca$_x$Se$_3$) progressively shifts $\mu_{b}$ from the conduction band to the valence band, thus changing as-grown \textit{n}-type Bi$_{2}$Se$_{3}$ into a \textit{p}-type degenerate semiconductor \cite{Cava1}. 
By exploiting the different  doping chemistry of Bi$_{2}$Se$_{3}$ (\textit{n}-type) and Bi$_{2}$Te$_{3}$ (\textit{p}-type), a better compensation was obtained in the Bi$_{2}$Se$_{2}$Te and Bi$_{2}$Te$_{2}$Se alloys \cite{Ando, Cava2}. In both cases a variable range hopping (VRH) behavior and, at low-T, a high resistivity (exceeding 1 $\Omega$cm) were observed \cite{Ando, Cava2}.

In this paper we present the first optical data of four topological insulators Bi$_{2-x}$Ca$_{x}$Se$_{3}$ ($x$ = 0, 0.0002), Bi$_{2}$Se$_{2}$Te and Bi$_{2}$Te$_{2}$Se from 5 to 300 K and from the sub-THz to the visible spectral range. The effects of the enhanced compensation are clearly visible in the far-infrared (FIR) spectra, through the suppression of the Drude term and the appearance of strong absorption peaks that we assign to electronic transitions among localized states, similar to those found in weakly doped semiconductors. Our data show that the electrodynamics of Bi$_2$Te$_2$Se, \textit{i.e.} the most compensated sample, is still affected by extrinsic doped charges, as therein the far-IR spectral weight is higher than the spectral weight associated with topological states by nearly two orders of magnitude.


\begin{figure}   \begin{center}  
\includegraphics[width=8cm]{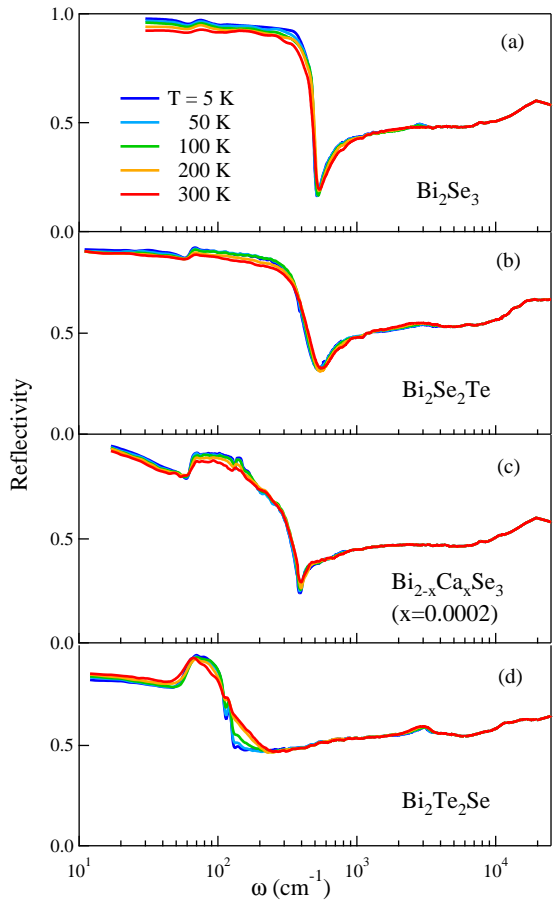}  
\caption{(Color online) Reflectivity of 
Bi$_{2}$Se$_{3}$ (a), Bi$_{1.9998}$Ca$_{0.0002}$Se$_{3}$ (b), Bi$_{2}$Se$_{2}$Te (c) and Bi$_{2}$SeTe$_{2}$ (d) from 10 to 24000 cm$^{-1}$ at different temperatures. The far-IR spectra in a), b) and c) are characterized by a free-carrier
plasma edge around 500 cm$^{-1}$  and 400 cm$^{-1}$, as well as phonon features $\alpha$ and $\beta$ at about 60 and 130 cm$^{-1}$ respectively. In d), due to a strong compensation (see text), the phonon absorptions can be clearer observed. In all spectra a weak bump develops at low-T around 3000 cm$^{-1}$ corresponding to the direct-gap transition. The triplet direct gap appears instead above 10000 cm$^{-1}$. } 
\label{Fig1}
\end{center}
\end{figure}                                                                                                                                                                                                                                        



Single crystals with $x$ = 0  were grown  by a modified Bridgeman method, those of Bi$_{2-x}$Ca$_x$Se$_3$ via a process of two-step melting \cite{Cava1}. Chemical compensation ($i.e.$ the insulating character) increases when passing from Bi$_2$Se$_3$ and Bi$_2$Se$_2$Te to Bi$_{2-x}$Ca$_x$Se$_3$ and Bi$_2$Te$_2$Se \cite{Cava3}. The basal ($ab$)-plane resistivity $\rho_{ab}(T)$ of the most compensated sample, Bi$_2$Te$_2$Se, shows an increasing (semiconducting) behavior down to about 50 K followed by a low-T regime in which resistivity saturates at values exceeding 1 $\Omega$cm. In this regime, surface charge-carrier mobility much higher than the bulk mobility has been detected \cite{Cava2, Cava3, Jia}. 
  
The reflectivity $R(\omega)$  of the four single crystals was measured at near-normal incidence with respect to the $ab$ basal plane from sub-THz to visible frequencies at temperatures ranging from 5 to 300 K, shortly after cleaving the sample. The reference was obtained by \textit{in-situ} evaporation of gold (silver) in the infrared (visible) range.  In the sub-THz region (below 30 cm$^{-1}$) we used Coherent Synchrotron Radiation (CSR) extracted from the electron storage ring BESSY II, working in the so-called \emph{low-$\alpha$} mode \cite{AboBakr-03}.
The real part of the optical conductivity $\sigma_{1}(\omega)$ was obtained from $R(\omega)$ via Kramers-Kronig  transformations,  after extrapolating $R(\omega)$ to zero frequency by Drude-Lorentz fits. The extrapolations to high frequency were based on data of Ref. \onlinecite{Greenaway}.

The reflectivity data for all samples are shown in Fig. \ref{Fig1}. The far-infrared spectra in Figs. \ref{Fig1}-a, -b are dominated by a free-carrier plasma edge around  500 cm$^{-1}$ , which confirms the picture of an extrinsic electrodynamics in these materials. In Bi$_{1.9998}$Ca$_{0.0002}$Se$_{3}$ (Fig. \ref{Fig1}-c), Ca doping shifts the plasma edge to about 400 cm$^{-1}$, while the strongest compensation is achieved in Bi$_{2}$Te$_{2}$Se (Fig. \ref{Fig1}-d) where unshielded phonons are well resolved at 60 ($\alpha$ mode) and 130 ($\beta$ mode) cm$^{-1}$. In all spectra a strong electronic absorption appears above 10000 cm$^{-1}$.


\begin{figure}   \begin{center}  
\includegraphics[width=8cm]{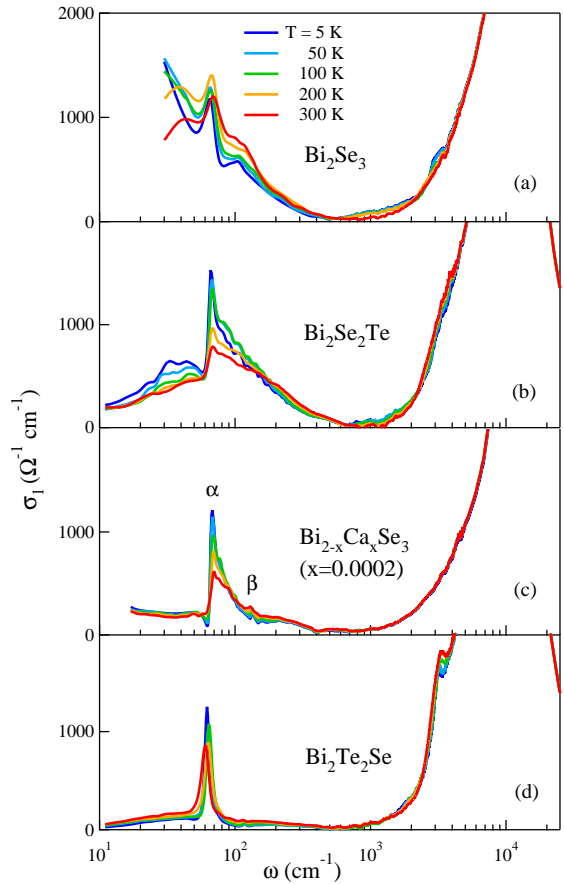}   
\caption{(Color online) Real part of the optical conductivity of 
Bi$_{2}$Se$_{3}$ (a), Bi$_{1.9998}$Ca$_{0.0002}$Se$_{3}$ (b), Bi$_{2}$Se$_{2}$Te (c) and Bi$_{2}$SeTe$_{2}$ (d) from 10 to 24000 cm$^{-1}$ at different temperatures. A broad minimum around 500 cm$^{-1}$ separates the high-frequency interband transitions from the low-energy excitations. The $\alpha$ and $\beta$ infrared-active phonon modes are indicated in panel (c).} 
\label{Fig2}
\end{center}
\end{figure}


The real part of the optical conductivity obtained from the $R(\omega)$ in Fig. \ref{Fig1} by Kramers-Kronig transformations is shown for the same temperatures and frequencies in Fig. \ref{Fig2}. 
The direct gap transition, which corresponds to a small bump around 3000 cm$^{-1}$ is barely visible due to its superposition with the huge triplet electronic excitation present above 10000 cm$^{-1}$. These interband features, which will be not the focus of this paper, have been discussed in detail in Ref. \onlinecite{Greenaway}.
Most of the effects induced by compensation appears below 500 cm$^{-1}$. In particular, the most extrinsic system Bi$_{2}$Se$_{3}$ (Fig. \ref{Fig2}-a), presents a Drude term superimposed to the $\alpha$ and $\beta$ phonon peaks, which both sharpen for decreasing T. A similar behavior has been observed by A. D. La Forge \textit{et al.} in Ref.  \onlinecite{Basov} on crystal with a charge-carrier density in the same range $\sim$10$^{18}$/cm$^3$.
The effect of compensation starts to be observable in Bi$_{2}$Se$_{2}$Te (Fig. \ref{Fig2}-b). Here, at variance with an appreciable $dc$ conductivity ($\sigma_{dc}\sim$200 $\Omega^{-1}$ cm$^{-1}$ \cite{Cava4}), most of the FIR spectral weight is located at finite frequency in the phonon spectral region. A further drastic reduction of the spectral weight is finally obtained in Bi$_{1.9998}$Ca$_{0.0002}$Se$_{3}$ and Bi$_{2}$SeTe$_{2}$ (Figs. \ref{Fig2}-c and -d).

The $\alpha$ phonon mode, both in Bi$_{2}$Se$_{2}$Te and Bi$_{1.9998}$Ca$_{0.0002}$Se$_{3}$, shows a Fano lineshape with a low frequency dip, more pronounced at low temperature. This suggests an interaction of this mode with an electronic continuum at lower frequency \cite{Damascelli}. The Fano shape is much less evident in Bi$_{2}$Te$_{2}$Se, where the $\alpha$ phonon shows a nearly Lorentzian shape at room $T$ and a weak low-frequency dip at low $T$. This indicates that the electronic continuum SW transfers from above to below the phonon frequency for increasing temperature \cite{lupi_PRB98}. At variance with previous samples, the $\alpha$ mode in Bi$_{2}$Se$_{3}$ shows instead a high-frequency dip at all T, confirming the observation in Ref. \onlinecite{Basov}. This behavior is in agreement with an electronic-continuum SW located in average at higher frequency with respect the $\alpha$-phonon characteristic frequency.

\begin{figure}   \begin{center}  
\includegraphics[width=8cm]{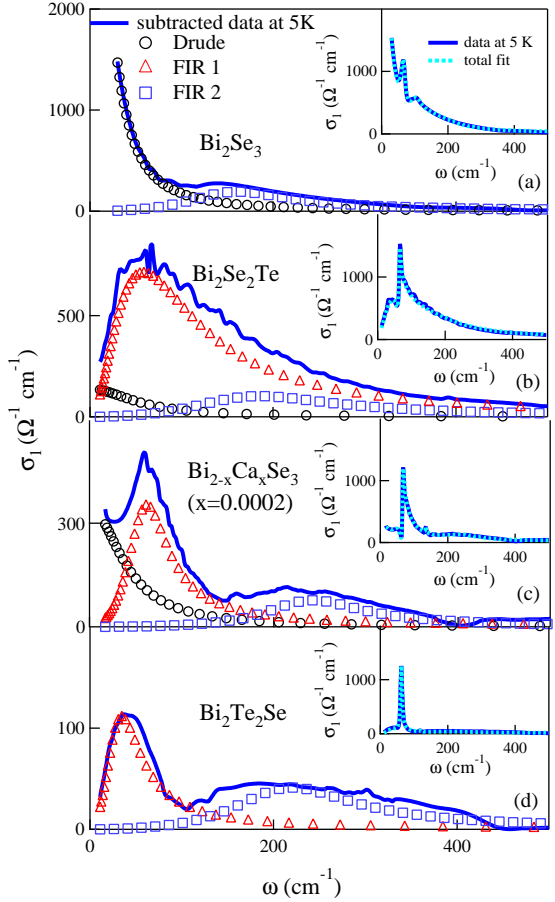}    
\caption{(Color online) Far-infrared optical conductivity at 5 K of the four Bi-based single crystals, after subtraction of both the interband and phonon contributions $via$ Drude-Lorentz fits.  The Drude term and the FIR contributions 1 and 2 are indicated by open circles, triangles, and squares, respectively. Note the different vertical scale in each panel.} 
\label{Fig3}
\end{center}
\end{figure}

In order to better follow the evolution of both the lattice and the electronic optical conductivity, we have fitted to $\sigma_{1}(\omega)$ a Drude-Lorentz (D-L) model where the $\alpha$ phonon mode is described in terms of the Fano shape \cite{Damascelli}. Examples of those fits (dotted lines) are shown in the insets of Fig. \ref{Fig3} at 5 K. 

The FIR $\sigma_{1}(\omega)$, as obtained after subtraction of both interband and phonon contributions, is  shown in the same Figure. The electronic conductivity of Bi$_{2}$Se$_{3}$ (Fig. \ref{Fig3}-a) can be described in terms of a Drude term (open circles) which narrows for decreasing $T$ in agreement with the metallic behavior of the resistivity \cite{Cava1}, and of a broad absorption centered around 150 cm$^{-1}$ (open squares). In Bi$_{2}$Se$_{2}$Te (Fig. \ref{Fig3}-b) most of the FIR spectral weight is located in a broad band centered around 100 cm$^{-1}$. This band has been modelled through two Lorentzian contributions,  located around 50 cm$^{-1}$ (open triangles, FIR1) and 200 cm$^{-1}$ (open squares, FIR2). In Bi$_{1.9998}$Ca$_{00002}$Se$_{3}$ the increased compensation results in an overall reduction of the FIR spectral weight. Moreover, the FIR absorption already observed in Bi$_{2}$Se$_{2}$Te (Fig. \ref{Fig3}-b) splits into two bands: a narrow absorption centered at about 50 cm$^{-1}$ and a broader one around 200 cm$^{-1}$. This double spectral structure is also achieved in Bi$_{2}$Te$_{2}$Se, where the low-frequency conductivity assumes a value comparable to the $\sigma_{dc}\sim$1 $\Omega^{-1}$ cm$^{-1}$ measured in crystals belonging to the same batch \cite{Cava2}. 


Similar low-frequency absorption bands in the optical conductivity have been observed in other, more conventional, doped semiconductors like Si:P \cite{Capizzi, Gaymann}. There, an insulator-to-metal transition (IMT) of Anderson type \cite{Mott}, can be observed for a charge-carrier density $n_{IMT}\sim$ 3.7 x 10$^{18}$/cm$^3$ \cite{Capizzi, Gaymann}. Deeply inside the insulating phase ($n < n_{IMT}$), narrow peaks can be observed in the far-IR corresponding to the hydrogen-like $1s\rightarrow np$ transitions of isolated P impurities. They are followed by a broad band at higher frequency, due to the transitions from the impurity bound states to the continuum. The narrow peaks broaden for increasing doping, giving rise to a low-frequency band which however remains distinguished from the higher-frequency absorption. When approaching the IMT the low-frequency band transforms into a Drude term, while the high-frequency absorption persists in the metallic phase, indicating that the IMT occurs in an impurity band \cite{Gaymann}. Therefore, in analogy with Si:P, the FIR2 band at 200 cm$^{-1}$ in Bi$_{1.9998}$Ca$_{0.0002}$Se$_{3}$ and Bi$_{2}$Te$_{2}$Se can be assigned to the transitions from the impurity bound states to the electronic continuum. The FIR2 band is also in very good agreement with the impurity ionization energy estimated from the T-dependence of resistivity and Hall data, namely $E_{i} \sim$ 30-40 meV \cite{Cava2, Ando, Bansal}. 
Instead, the low-frequency FIR1 band, clearly resolved both in Bi$_{1.9998}$Ca$_{0.0002}$Se$_{3}$ and Bi$_{2}$Te$_{2}$Se (Fig. \ref{Fig3}-c and -d, respectively) can be associated with the hydrogen-like $1s\rightarrow np$ transitions, broadened by the inhomogeneous environment of the impurities and/or by their interactions. The spectra in Fig.  \ref{Fig3} clearly show that, even in the most compensated topological insulator here measured, $i.e.$ Bi$_{2}$Te$_{2}$Se, the far-IR conductivity is still affected by extrinsic charge-carriers injected into the system by non-stoichiometry and doping. 

\begin{figure}   \begin{center}  
\includegraphics[width=8cm]{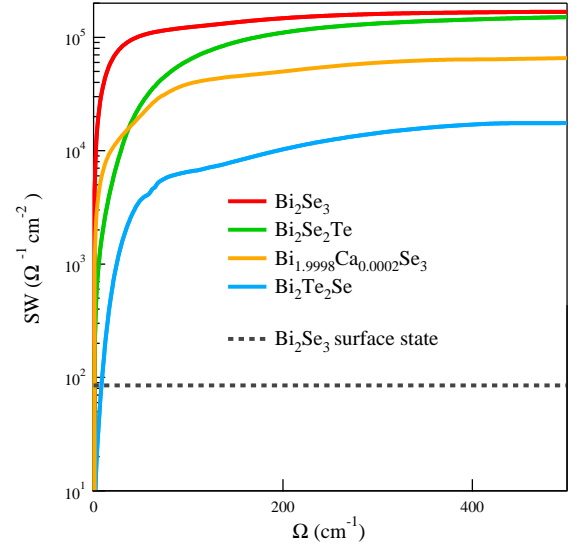}    
\caption{(Color online) FIR spectral weight of all samples calculated at 5 K by Eq. \ref{Eq1}. The dashed curve represents the contribution from the metallic surface carriers as obtained from data of Refs. \onlinecite{Armitage1}.}
\label{Fig4}
\end{center}
\end{figure}

A more quantitative comparison between the charge density expected for the topological surface states and that provided by extrinsic charge carriers can be obtained by calculating the optical spectral weight $SW{(\Omega)}$:

\begin{center}
\begin{equation}
SW{(\Omega})=\int_0^{\Omega}\sigma_{1}^{sub}(\omega)\mathrm{d}\omega
\label{Eq1}
\end{equation}
\end{center}

\noindent

where $\Omega$=500 cm$^{-1}$, a cut-off frequency which well separates (see Fig.\ref{Fig3}), the low-frequency excitations from the interband transitions. 
In Fig. \ref{Fig4} one observes a $SW$ decrease of one order of magnitude when passing from Bi$_{2}$Se$_{3}$ to Bi$_{2}$Te$_{2}$Se which confirms the drastic effect of chemical compensation. From the $SW$ of Bi$_{2}$Te$_{2}$Se and assuming a carrier mass m$^{*}\sim0.01m_{e}$ \cite{Cava1}, one can estimate a 3D charge density $n\sim10^{17}$/cm$^3$. This value fits very well the interval 5 x 10$^{16}$-2 x 10$^{17}$/cm$^3$ obtained from transport measurements in crystals of the same batch \cite{Jia}.

Recently, the THz conductivity of Molecular Beam Epitaxy (MBE) Bi$_{2}$Se$_{3}$ films has been measured by a Time Domain Spectrometer from 200 GHz to about 2 THz \cite{Armitage1, Armitage2}. The optical conductivity in this spectral region, can be satisfactorily fits through a 2D Drude term plus the (3D) $\alpha$ phonon mode. The 2D Drude contribution has been associated to topological surface states \cite{Armitage1, Armitage2}.
Therefore their spectral-weight can be estimated from the plasma frequency ($\omega_{SSp}$) as obtained from the Drude fit in Refs. \onlinecite{Armitage1}: $SW_{SS}=\omega_{SSp}^2/8$. This value has been plotted in Fig. \ref{Fig4} through a dashed line. The actual $SW$ in Bi$_{2}$Te$_{2}$Se is still nearly two orders of magnitude higher than that expected from the topological surface states. This result demonstrates that the low-energy electrodynamics in 3D single crystals of topological insulators, even at the highest degree of compensation presently achieved, is still affected by extrinsic charge excitations.


A further optimization in the crystal compensation is expected to provide crystals in which bulk techniques like infrared spectroscopy should be able to observe the intrinsic optical properties due to 2D surface metallic states.

We acknowledge the Helmholtz-Zentrum Berlin -  Electron storage ring BESSY 
II for provision of synchrotron radiation at beamline IRIS. The research leading to these results has received funding from the 
European Community's Seventh Framework Programme (FP7/2007-2013) under grant 
agreement n.226716. The crystal growth  was supported by the US national Science Foundation, grant DMR-0819860.



\end{document}